\documentclass[aps,twocolumn,showpacs,floatfix]{revtex4}
\usepackage{graphicx}

\let\oldmarginpar\marginpar
\renewcommand\marginpar[1]{\-\oldmarginpar[\raggedleft\small\sf #1]{\raggedright\small\sf #1}}

\newcommand{\FigWidth}{\columnwidth}

\begin{document}

\title{The spin resonance and high frequency optical properties
 of the cuprates.}
   
\author{Ar. Abanov$^1$ and Andrey V. Chubukov$^2$}
\affiliation{$^1$Los Alamos National Laboratory, MS 262B, Los Alamos, NM 87545}

\affiliation{$^2$Department of Physics, University of Wisconsin, Madison, WI 53706}
\date{\today}

\begin{abstract}
We argue that recently observed superconductivity-induced 
blue shift 
 of the plasma frequency $\delta \omega_{pl}$  in 
$Bi_2Sr_2CaCu_2O_{8+\delta}$~\cite{vdm}
 is related to the change in the integrated dynamical structure factor 
associated with the development of the spin resonance below $T_c$. 
We show that the magnitude of $\delta \omega_{pl}$ 
is consistent with the small integrated spectral weight of the resonance, and
 its temperature dependences closely  follow that of the spin resonance peak.
\end{abstract}
\maketitle

The importance of the resonance spin mode for
 the physics of the cuprates continue
 to be the subject of intensive debate. In a generic superconductor, 
the pairing of fermions drastically reduces the damping of collective 
spin degrees of freedom at energies below $2\Delta$. 
For a $d-$wave superconductor, 
 the residual interaction between spin fluctuations and fermions
 gives rise to the additional effect -- 
 the development of the exciton mode below $2\Delta$ (see e.g. Ref. \cite{mike_1} and references therein). 
This mode exists for bosonic momenta near $(\pi,\pi)$ and is commonly called 
the ``spin resonance''. It has been observed in 
three different families of high-$T_c$ superconductors~
\cite{rosa,fong,dai,bi,keimer}. 
This mode is not a ``glue'' to superconductivity as it
 emerges only in the superconducting state (more precisely, below the 
 pseudogap temperature), but it 
affects electronic properties of the cuprates in the superconducting state. 

Much of recent works on the effect of the spin resonance on electrons was 
 concentrated on whether the interaction with the resonance
is capable to explain 
  experimentally detected {\it low-energy} features in the fermionic 
spectral function, 
 tunneling density of states 
and optical conductivity~\cite{mike_1,acs}. 
 An example of this behavior is the peak/dip/hump structure of
 the spectral function~\cite{ARPES_pdh}.

The spin-resonance scenario for the cuprates was recently 
questioned~\cite{kee} on the basis that the
measured  spectral weigh of the resonance is only $1-2$ percent  
of the total magnetic spectral weight. 
Ar. Abanov et al., however,  argued~ \cite{reply} that 
this smallness merely reflects the fact that the 
resonance exists only in a narrow momentum range near ${\bf Q} = 
(\pi/a, \pi/a)$ [roughly, between ${\bf Q}$ and $0.8 {\bf Q}$], 
while at momenta where the 
resonance does exist, it strongly couples to fermions. 
They further argued that at small $\omega \sim 2\Delta$,
 the fermionic self-energy
 change between normal and superconducting states chiefly comes from
 spin fluctuations with energies comparable to $\omega$, and 
 from spin  momenta $|{\bf q}-{\bf Q}| \sim \omega/v_F$, where $v_F$ is
 the bare Fermi velocity. As $v_F \sim 1 eV*a$ is large,   
 typical spin momenta  for $\omega \leq 100 meV$
 are well within the range where the resonance 
does exist, i.e., the smallness of the total spectral weight of 
the resonance does not matter. 

The subject of  the present communication is the analysis of the 
 possible role 
of the spin resonance in the 
observed changes between normal and superconducting state 
in the optical data at high frequencies, $\omega \sim 0.5-1eV$. 
Recently, Molegraaf et al. 
reported the results of their ellipsometry 
measurements on optimally doped and underdoped  
$Bi_2Sr_2CaCu_2O_{8+\delta}$ with $T_{c}=88K$ and $T_{c}=66K$ 
respectively~\cite{vdm}. 
They observed that  in the normal state, 
 the in-plane plasma frequency increases
as $T^2$ with decreasing $T$,
 however below $T_c$ it increases faster such that the actual 
value of $\omega_{pl} (T=0)$ is larger than the extrapolation from 
the normal state. The effect  is very small: at optimal doping  
$\delta \omega_{pl} \sim 10 cm^{-1}$  is only 
$1.3 \times 10^{-3}$ of the plasma frequency 
 $\omega_{pl} \sim 7600 cm^{-1} \approx 1 eV$, 
but detectable by the ellipsometry technique.    
 
The change of the plasma frequency in a superconductor
 is related to the change of the fermionic 
self-energy between 
superconducting and normal states (see Eq. (\ref{0}) below).
 Conventional wisdom holds that at $\omega \sim \omega_{pl}$, 
 which well exceed
 the magnetic bandwidth, interaction with low-energy spin fluctuations 
is not the dominant mechanism for the fermionic self-energy. 
We argue, however, that while this is generally true 
for $\Sigma (\omega)$ itself, 
 $\delta \omega_{pl}$ scales with $\delta \Sigma (\omega) = \Sigma_{sc} (\omega) - \Sigma_n (\omega)$, and the latter  
comes from  frequencies comparable to the superconducting 
gap and can be captured within the low-energy, spin-fluctuation theory.
  We will see, however, that at high $\omega$, the concept that intermediate fermions and bosons have equal energies fails,
 and a fermion with energy $1 eV$ interacts with the whole band of 
magnetic fluctuations. As a result, 
  $\delta \Sigma (\omega)$ scales with the 
 integrated magnetic spectral weight, and 
 the smallness of the  spectral weight transfer into the resonance
 becomes crucial.
We argue that the value of the observed 
  shift of the plasma frequency is
 consistent with the fact that only $1-2\%$ of the 
magnetic spectral weight is transferred into the resonance. 

Our reasoning is the following. At plasma frequency, the real part of the
 dielectric function $\epsilon (\omega)$ changes sign. 
The dielectric function obeys 
$\epsilon (\omega) = \epsilon (\infty) + 4 \pi i \sigma (\omega)/\omega$, where 
$\sigma (\omega)$ is the optical conductivity. By Kubo formula, 
$\sigma (\omega) = ((\omega^0_{pl})^{2}/(4 \pi)) Re [\Pi (\omega)/(-i \omega)]$ where
 $\Pi (\omega)$ is fully renormalized current-current polarization 
operator and $(\omega^{0}_{pl})^{2} = 4 \pi n e^2/m$
 is the bare plasma frequency.
The plasma frequency is then the solution of  
\begin{equation}
\omega_{pl} = \frac{\omega^0_{pl}}{\sqrt{\epsilon (\infty)}}~\sqrt{Re \Pi (\omega_{pl})}
\label{01}
\end{equation}
To zero-order approximation, $\Pi (\omega_{pl}) =1$, 
i.e., $\omega_{pl} =\omega^0_{pl}/\sqrt{\epsilon (\infty)}$. 
However, at any finite frequency $1-Re \Pi (\omega)$ is still finite, and 
hence $\omega_{pl}$ is sensitive to the
change of the polarization operator upon entering the superconducting state.
 This change of $\omega_{pl}$  is  
small as superconductivity 
 mostly affects the form of $\Pi (\omega)$ at frequencies comparable 
to the superconducting gap $\Delta \sim 0.04 \omega_{pl}$.
 
At high frequencies,  $\omega \sim 1eV$, 
 normal and anomalous fermionic self-energies  
$\Sigma ({\bf k}, \omega) \approx \Sigma (\omega)$ and 
$\Phi ({\bf k}, \omega)$ 
 are both small compared to $\omega$,
  and to the leading order in the self-energy, the current-current correlator 
 both in the normal and  superconducting states is given by
\begin{equation}
\Pi (\Omega) = \int_0^\Omega \frac{d\omega}{\Omega + \Sigma (\omega) + 
\Sigma (\Omega -\omega)}
\label{pi}
\end{equation}
Substituting this expression into (\ref{01}) we find that 
 the superconductivity-induced change in the plasma frequency  at $T=0$ is
 related to the difference between  superconducting and normal 
fermionic self-energies $\delta \Sigma (\omega) = \Sigma^{sc} (\omega) - \Sigma^{n} (\omega)$ as
\begin{eqnarray}
&&\frac{\delta \omega_{pl}}{\omega_{pl}} = - 
\frac{1}{2 \Pi^\prime_n (\omega_{pl})}~\nonumber \\ 
\!\!\!\!\!\times Re&&\!\!\!\!\! \int_0^{\omega_{pl}} \!\!\!\!\!
d\omega~\frac{\delta \Sigma (\omega) + 
\delta \Sigma (\omega_{pl} - \omega)}{\left[\omega_{pl} + \omega Z_\omega +
(\omega_{pl} -\omega)Z_{\omega_{pl} -\omega}\right]^2} 
\label{0}
 \end{eqnarray}
where $
\delta \omega_{pl} = \omega^{sc}_{pl} - \omega^n_{pl}$, 
$\Pi^{\prime}_n (\omega)$
is the real part of the polarization operator 
in the normal state, and $Z_\omega = 1 + \Sigma_n (\omega)/\omega$  
is inverse quasiparticle residue in the normal state. 
By all accounts, at $\omega \sim \omega_{pl}$, $\Sigma (\omega) 
\ll  \omega$, i.e., $Z_{\omega} \approx 1$. Hence, once
 the integral in (\ref{0}) 
is dominated by frequencies  where  either $\omega$ or 
$\omega_{pl} - \omega$ are near $\omega_{pl}$ (as we later verify),
 the precise form of the fermionic $Z(\omega)$ at high frequencies does not 
matter, and   $\delta \omega_{pl}/\omega_{pl} \approx 
\delta \Sigma (\omega_{pl})/(2 \omega_{pl})$.
 
The computation of $\delta \omega_{pl}$ therefore reduces 
to the computation of the self-energy difference between normal and 
superconducting states. 
We present the result for $\delta \Sigma (\omega)$ now and discuss its 
derivation later. We found that, within the spin-fluctuation scenario,  
 there are two distinct frequency regimes depending, roughly,
 on whether or not $\omega$ exceeds the magnetic bandwidth. At small
 frequencies,  the Eliashberg  approximation is valid, and 
$\delta \Sigma (\omega)$ is positive  (see Fig. \ref{fig5}).
 In this regime,  internal fermions and bosons 
in the self-energy diagram have comparable energies, i.e., 
a fermion is interacting only with spin fluctuations very near ${\bf Q}$. 
 We verified that
 if this behavior extended up to $\omega \sim \omega_{pl}$, $\delta \omega_{pl}$ would be negative, in disagreement with the data.     

At larger frequencies, however, Eliashberg approximation becomes invalid,
 and a novel (anti-Eliashberg) approximation has to be used. 
In this approximation, we obtained that the change of the
fermionic self-energy is proportional to
the {\it integrated} change of the dynamical spin structure factor
\begin{equation}
\delta \Sigma (\omega) \approx -\frac{3g^2}{Z_{\omega} \omega} 
\int \frac{d \Omega d^2 q}{8\pi^3} \sum_i \delta 
S_{i} ({\bf q}, \Omega) F({\bf q}) 
\label{1}
\end{equation}
where $g$ is the spin-fermion coupling constant 
 estimated to be $g \sim 0.7 eV$~\cite{acs,mike_1,reply}, and 
$\delta S_{i} ({\bf q}, \Omega) =  
S^{sc}_{i} ({\bf q}, \Omega) - S^{n}_{i} ({\bf q}, \Omega)$ is the change in the  dynamical structure  
factor $S({\bf q}, \Omega) = \chi^{\prime \prime} 
({\bf q}, \Omega) (1 + \coth (\omega/2T))$ in even ($i=1$) and odd ($i=2$) channels. 
The factor $F({\bf q})$  decreases  away from ${\bf Q}$ but 
 can be safely approximated by $F ({\bf q}) =1$ 
in the narrow momentum range where the resonance is experimentally detectable.
The integrated magnetic spectral weight near $(\pi,\pi)$ is larger
 in the superconducting state~\cite{fong,dai}, hence at high frequencies 
$\Sigma^\prime (\omega)$ is {\it smaller}  in a superconductor. Using this
$\delta \Sigma (\omega)$ we find that $\delta \omega_{pl}$ is positive, in
 agreement with \cite{vdm}.

The crossover between the two regimes occurs at optimal doping at 
$\omega = \omega_0 
\sim 300 meV$, see Fig. \ref{fig:sigma}. Theoretically, we 
 use one-band model to get Eq. \ref{1}, i.e., we assume 
 that there is a frequency range above $\omega_0$, 
where anti-Eliashberg approximation
 is valid, but interband transitions still can be neglected. The applicability
 of this approximation has to be verified by comparing the results of one-band
 analysis with the data.   
 
 The momentum and frequency  integral in the r.h.s. of  (\ref{1}) yields
$B S (S+1)/3$ where $B$ is the percentage of the spectral weight 
redistributed below $T_c$.   
Only the odd channel 
contributes to $\delta S ({\bf q}, \Omega)$ and  $1-2\%$ 
of the spectral weight 
from this channel is redistributed \cite{fong}, i.e., 
$B \sim 0.005-0.01$. 
Substituting $\delta \Sigma (\omega) \approx -3g^2 B/(4\omega)$  
into the expression for $\delta \omega_{pl}$ we obtain 
$\delta \omega_{pl}/\omega_{pl} \sim 
(1-2) \times 10^{-3}$, in near perfect agreement with the experimental result
 $1.3 \times 10^{-3}$~\cite{vdm}. 
We emphasize that the 
agreement is entirely due to the fact that the 
integrated weight of the resonance is very small. 
If it wasn't, the blue shift of the plasma frequency would be much larger.
To verify the prefactor, we went beyond estimates and  
  evaluated the full integral in (\ref{0}) using the normal state 
expression for $Z_{\omega}$~\cite{acs}:
 $Z_\omega  \approx 1 + (i {\bar \omega} /\omega)^{1/2}$ 
where ${\bar \omega} \sim 0.35 g$.
 We obtained almost the same result
  for $\delta \omega_{pl}$ as above -- the extra  prefactor is 
 $0.97 \approx 1$. 

\begin{figure}
\includegraphics[clip=true,width=\FigWidth,height=1.9in]{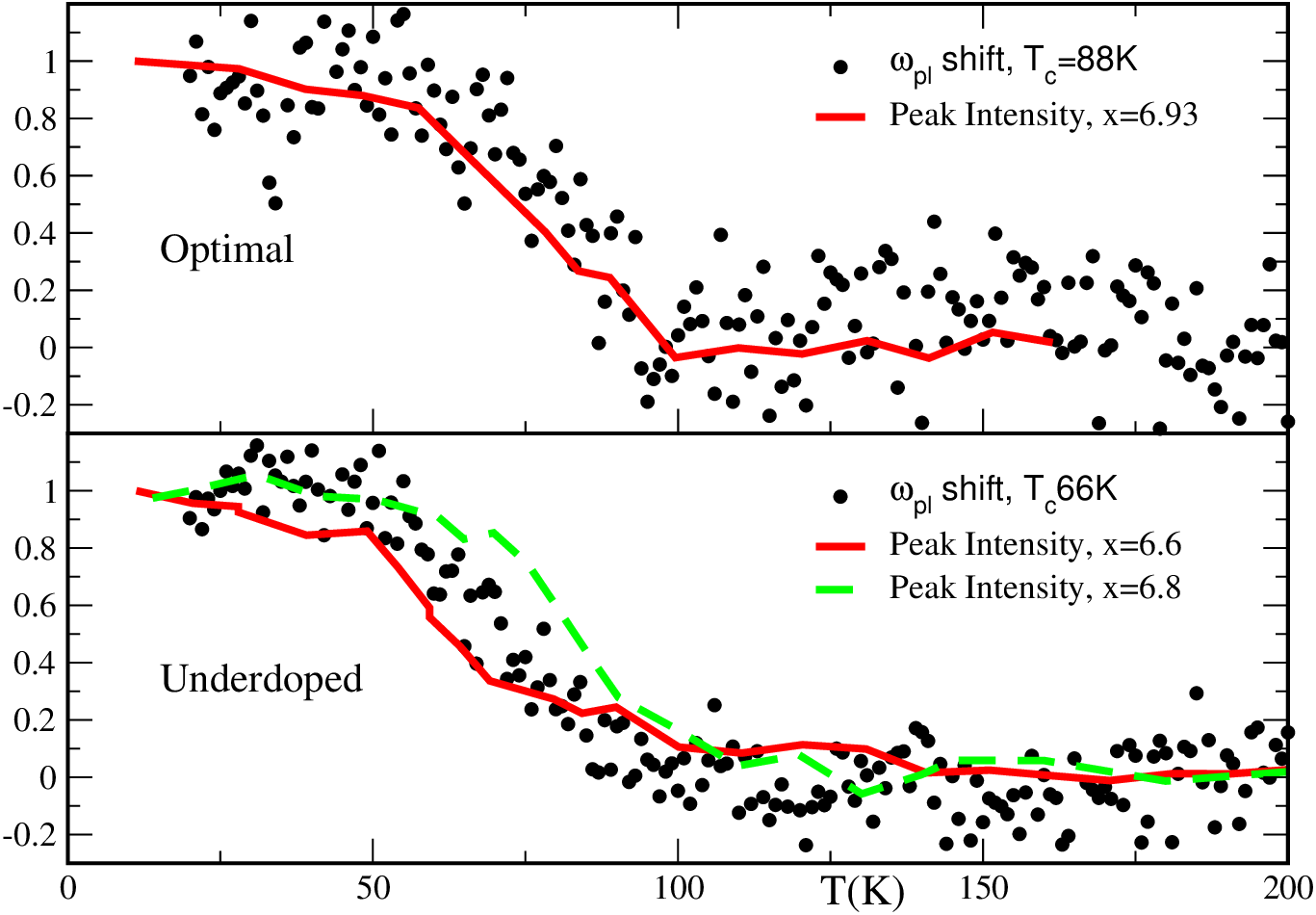}
\caption{\label{fig:plasma} The temperature dependence of the 
superconductivity-induced  change of the plasma frequency in $Bi2212$
\cite{vdm}
 vs the change in the  resonance peak intensity for $YBCO$ \cite{dai}.
 Upper panel - optimal doping, lower panel - underdoped samples.
 We plot $YBCO$ data for two different underdoped samples for comparison. }
\end{figure}

\begin{figure}
\includegraphics[clip=true,width=\FigWidth,height=1.1in]{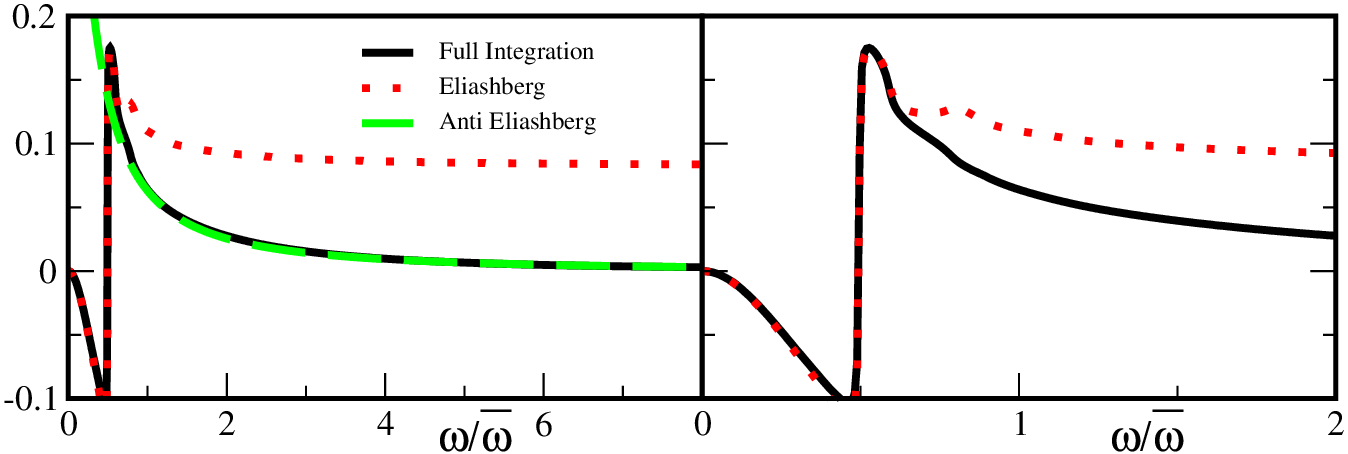}
\caption{\label{fig:sigma} 
Right panel - the difference in the electronic self-energy between normal and superconducting
states calculated explicitly and using Eliashberg and anti-Eliashberg 
 approximations. The frequency is measured in units of
 ${\bar \omega} \sim 0.35 g \sim 250 meV$. Clearly, the 
anti-Eliashberg approximation is 
 much better at large frequencies.
  Left panel - the 
low-frequency region, where the Eliashberg approximation is valid.}
\end{figure}

\begin{figure}
\includegraphics[clip=true,width=\FigWidth,height=1.05in]{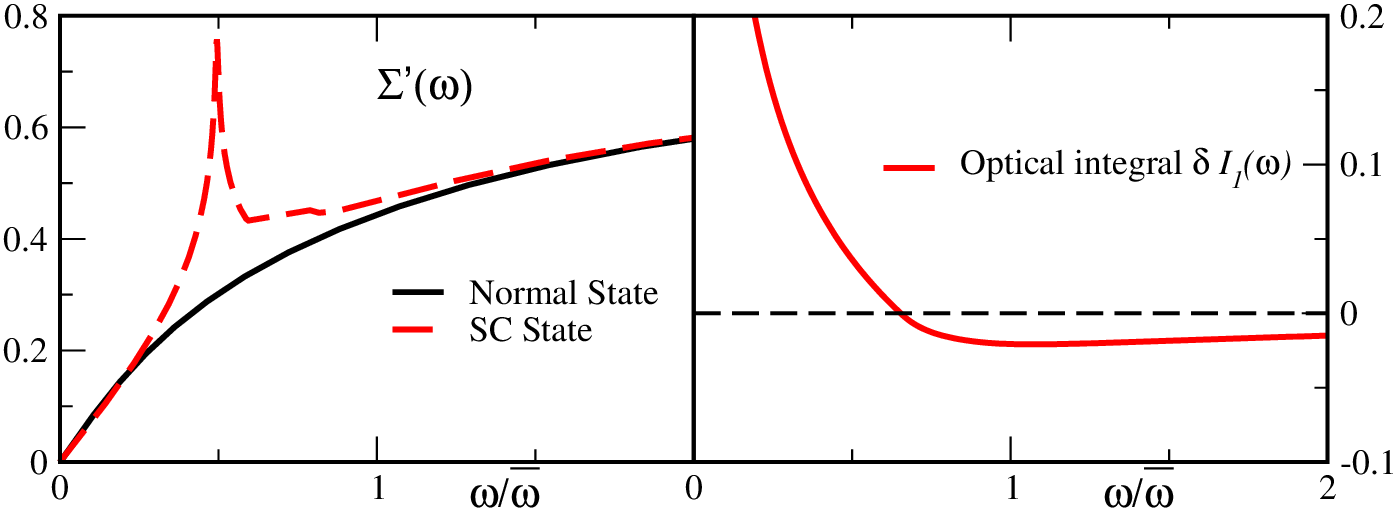}
\caption{\label{fig5} 
The results of Eliashberg calculations for coupling $\lambda =1$. 
Left panel - self-energy, right panel - the optical integral $\delta I(\omega)$ (including the  condensate piece). 
 For $\lambda =2$, $\delta I(\omega)$ changes 
sign at a larger $\omega \sim 2 {\bar \omega}$.}
\end{figure}

In Fig.\ref{fig:plasma} 
  we plot  the temperature dependence of 
$\delta \omega_{pl}(T)$  together with the temperature dependence 
 of the resonance peak intensity~\cite{dai}.
The data for the fully integrated intensity are available for fewer
 temperatures and show roughly the same $T$ dependence~\cite{dai}. 
 We see that the $T$  dependences of $\delta \omega_{pl}$ and of the 
 resonance peak  follow each other as it should be according to 
 Eq. (\ref{1}). 

We also searched for the explanation of 
 the $T^2$ temperature dependences of $\omega_{pl}$ found by 
Molegraaf et al. in the normal state. 
 The analysis of the neutron data~\cite{fong}
 shows that the dominant source of the temperature dependence of the  
spectral function is the $T$ dependence of $\chi^{\prime \prime} (\omega)$,
 while the one from $(1 + \coth \omega/(2T))$ is much weaker. 
 Indeed, the neutron data presented for $\omega = 65 meV$~\cite{fong} 
 show that  
the local $\chi^{\prime \prime} (\omega)$ decreases by 
more than the factor of $2$ between $100$ and $200K$, when the
 thermal factor $\coth (\omega/2T) \approx 1$ for both temperatures. 
It is therefore likely that the $T$ dependence in the normal state  
 originates in thermal corrections to the parameters of the dynamical 
  spin susceptibility (i.e., to the magnetic correlation length).
 Theoretically, 
these corrections form regular series in $(T/E_F)^2$~\cite{acs}, 
 hence the temperature dependence
 of  $\omega_{pl}$ must  be $T^2$, unless superconductivity interferes.

We now turn to the conductivity.
Molegraaf et al. reported that  
 the Drude optical spectral weight integrated 
 up to a frequency of $10 000 cm^{-1}$ (including 
the condensate contribution) 
increases below $T_c$, and argued
 that this increase is compensated by the 
 decrease of the optical spectral weight integrated between 
$10 000 cm^{-1}$ and $20 000 cm^{-1}$, where interband transitions are overly relevant.   This, they argued, implies that the condensate contribution 
to $\sigma_{1} (\omega )$ is not 
compensated by the reduction of the conductivity in the superconducting state
 up to frequencies when one-band description fails.

Our results are inconsistent with this claim.
 The optical integral  
$\delta I (\Omega) =  \hbar^2 \int^{\Omega}_0 \delta 
\sigma (\omega) d \omega$ 
for $\Omega \sim 1eV$ covers the range  of frequencies
 where the Eliashberg approximation is valid ($\delta I_1$) , and the 
range where Eq. (\ref{1})
 is valid ($\delta I_2$). Using Eq. (\ref{1}), we can  compute  
 the partial integral $\delta I_2 (\Omega)=  \hbar^2 \int^{\Omega}_{\omega_0}
 \delta \sigma (\omega) d \omega$ where, we remind, $\omega_0$ is the 
lower boundary for the applicability of Eq. (\ref{1}). 
The approximations for $\sigma$ 
 are less robust than for $\omega_{pl}$ as for the conductivity we 
need to know $Im Z_{\omega}$ at high frequencies, 
where the use of the low-energy theory is questionable.  
Still, at high frequencies, 
$Im [1/Z(\omega)] \approx - Im \Sigma (\omega)/\omega$ is negative, 
 hence, according to (\ref{1}),  the high-frequency conductivity 
 $\sigma (\omega) \sim 1/\omega^2 \tau (\omega) 
\propto \Sigma^{\prime \prime} (\omega)$ is 
larger in the superconducting state than in the normal 
state extrapolated to $T=0$. 
 Accordingly,  $\delta I_2 (\Omega)$ is positive. 
 Substituting Eqs. (\ref{pi}) and (\ref{1})
 into the Kubo formula, and evaluating the integrals with the same $Z(\omega)$ as above, we obtain 
$ \delta I_2 (\Omega) 
\approx (2-4)\times 10^{-3} (eV)^2$ for $\Omega \sim 1 eV$. 

The  integral for $ \delta I_2 (\Omega)$
   converges at the upper limit; formally taking 
$\Omega = \infty$  yields almost the same $\delta I_2$.
As the $f-$sum rule must be satisfied within the one-band model (if 
 the bandwidth is formally set to infinity), 
the value $-\delta I_2 (\infty)$ gives the estimate of 
the optical integral $\delta I_1 (\omega_0)$ over frequencies 
where the Eliashberg theory is valid. 
 From our consideration, this integral is negative, i.e., the 
condensate contribution is overcompensated by the reduction of 
the conductivity in the superconducting state  below $\omega_0$~\cite{comm3}. 
As an independent check, we computed optical integral 
$\delta I_1$ within Eliashberg 
theory~\cite{ac_2}, and indeed obtained
 that $\delta I_1$ changes sign at $\omega \sim 300 meV$ at optimal doping 
(see Fig.\ref{fig5}). The  accuracy of our  
 Eliashberg
 calculations is not sufficient  to compare the values of the 
two contributions to $\delta I$. 

Our results therefore indicate that at $T=0$, 
the full optical integral is exchausted 
 at frequencies smaller than the bandwidth.
 The optical integral evaluated over larger frequencies is larger 
 at $T=0$  than in the normal state extrapolated to $T=0$. 
 The crossover frequency $\omega_0$ somewhat increases with underdoping 
(see caption for Fig.\ref{fig5}) but theoretically it 
 still remains $ O({\bar \omega})$ even in strongly underdoped 
 materials~\cite{homes}. 
Note that this does not contradict the idea that 
 that superconductivity is driven by the decrease of the 
 kinetic energy~\cite{norman_pepin}, as within the same approach, the decrease 
 of the  kinetic energy also comes from frequencies $(1-2) 
{\bar \omega}$~\cite{rob}.    

We now describe the calculation of the self-energy, 
Eq. (\ref{1}) in some detail.
We assume that the  fermionic self-energy predominantly comes from
 the fermion-fermion interaction in the spin channel and  can be 
  viewed as being mediated by  
 spin collective modes with momenta near $(\pi,\pi)$. 
Quite generally, the imaginary part of the 
fermionic self-energy  is given by 
\begin{eqnarray}
\Sigma^{\prime \prime} ({\bf k}, \omega) &=& 
\frac{3}{8 \pi^3} \int {\tilde g}^2 
\chi^{\prime \prime} ({\bf q}, \Omega) 
~G^{\prime \prime} ({\bf k}+{\bf q}, \omega + \Omega)~ \nonumber \\
&&\left[\tanh {\frac{\omega - \Omega}{2T}}+ \coth{\frac{\Omega}{2T}}\right]
d\Omega d^{2}q
\label{2}
\end{eqnarray}
where ${\tilde g}$ is the fully renormalized vertex, 
$\chi ({\bf q}, \Omega)$ is the propagator of the collective mode, 
and $G ({\bf k}+{\bf q}, \omega + \Omega)$ is the full fermionic 
Green's function. 

At small frequencies, the 
 physics that allows one to neglect  vertex 
 corrections (i.e., set ${\tilde g} = g$), is the 
fact that at strong coupling, spin fluctuations are 
overdamped in the normal state and  
are slow modes compared to electrons. Hence, an 
effective Migdal theorem is valid~\cite{acs}. By the same reason, 
 the momentum integration in Eqn (\ref{2}) is 
 factorized -- the integration transverse to the Fermi surface involves 
only fast fermion, while the integration along the Fermi surface is over 
 a slow bosonic momentum.
This computational
 procedure is  called Eliashberg approximation.
It describes the jump in $\Sigma^{\prime \prime}$ at $\Delta + \omega_{res}$ 
 that is the key element of the peak/dip/hump behavior~\cite{mike_1,acs}.
However, for $\delta \Sigma (\omega)$ this 
approximation is valid only as long as external 
fermionic frequency $\omega$ is smaller than a typical frequency at which 
 the momentum integral of $\delta \chi^{\prime \prime} (\Omega)$ converges.
At strong coupling, this generally happens at $\Omega$ comparable to the
 effective bosonic bandwidth ${\bar \omega}$ defined such that in the normal state, $\Sigma (\omega)$ becomes less than $\omega$ at $\omega > {\bar \omega}$, i,e., spin-fluctuation scattering becomes ineffective.  When spin-fermion coupling $g$ is less than the fermionic bandwidth $W$, 
${\bar \omega} \sim 0.35 {\bar g}$, in the opposite limit 
 ${\bar \omega} \sim W^2/g \sim J$.  Note that within the same model, 
$\Delta \sim 0.3 {\bar \omega}$~\cite{acs}.  

When $\omega \gg  {\bar \omega}$, and $k \approx k_F$,
 $G^{\prime \prime} ({\bf k}+{\bf q}, \omega + \Omega) \approx
 Im (\omega Z_{\omega })^{-1}$, 
and it can be 
taken out of the integral. To the same accuracy,
 the $T$ dependent factor in (\ref{2}) is approximated 
by $1 + \coth( \Omega/2T)$. We then obtain:
\begin{eqnarray}
&&Im \left[\delta \Sigma (\omega) \right] =
-3g^2~ Im \left[\frac{1}{Z_{\omega} \omega}\right]~\times \nonumber \\
&& \int \frac{d \Omega d^2 q}{8\pi^3} \sum_i \delta 
S_{i} ({\bf q}, \Omega) F({\bf q})
\label{1a}
\end{eqnarray}
where $F({\bf q})$ subject to $F({\bf Q}) =1$ decreases at 
$|{\bf Q}-{\bf q}| = O(|{\bf Q}|)$ and reflects the fact that
the spin-fermion model is only valid for bosonic momenta near $(\pi,\pi)$.
As the Kramers-Kronig transform of (\ref{1a}) is infrared convergent, 
 the full self-energy is 
 given by (\ref{1}).

We find therefore that at 
 high frequencies $\omega \gg {\bar \omega}$, 
 the correct computational procedure for $\delta  \Sigma (\omega)$ 
is opposite to the Eliashberg approximation -- instead of 
factorizing the momentum integral, one  can neglect
 the momentum  dependence in the Green's function and perform
 the full 2D momentum integration over the bosonic momenta.   
We verified that in this, anti-Eliashberg approximation, vertex corrections are  again small, this time in ${\bar \omega}/\omega$, such that ${\tilde g} = g$  in  Eq. (\ref{2}). 

Obviously, there should  be a crossover between Eliashberg and 
anti-Eliashberg approximations as frequency increases. To understand where
 it is located, we evaluated $\delta\Sigma^{\prime \prime} (\omega)$
 explicitly, using the normal and superconducting forms of the dynamical spin 
susceptibility obtained earlier~\cite{acs}, and compared the full result 
with the two approximate forms.
The results are presented in Fig. \ref{fig:sigma}. We see that  
for $\omega \leq {\bar \omega}$, Eliashberg approximation is much closer to 
 the full result. However, for $\omega > (1-2)\bar{\omega}\approx 250-500 meV$ 
 the  Eliashberg approximation is well off,
 while the 
 anti-Eliashberg approximation is rather close to the full expression. 
This justifies our use of Eqn. (\ref{2})
 for optical properties above $500 meV$.

 To summarize, in this paper we argued that
the superconductivity-induced 
blue shift of the plasma frequency,
 detected in the ellipsometry studies,  can be explained within the magnetic 
scenario for the cuprates. We found that $\delta \omega_{pl}$  scales with
 the change of the integrated  magnetic spectral 
weight $\delta S({\bf q}, \Omega)$. 
The magnitude of $\delta \omega_{pl}/\omega_{pl}$ is small ($\sim 10^{-3}$)
 as the integrated $\delta S({\bf q}, \Omega)$   accounts for only 
 a  small fraction  of the total spectral weight. 
We also predict that the optical integral converges below $0.5 eV$.
 Careful measurements of this 
integral  should either confirm or disprove our claim.

We are thankful to D. Basov, G. Blumberg,
  C. Homes, B. Keimer,
 H.J.A. Molegraaf, M. Norman,  and D. van der Marel for useful discussions and
 to P. Dai and H.J.A. Molegraaf for providing us with  their  data.
The research was supported by
NSF DMR 0240238 (A. Ch.) and by 
Los Alamos National Laboratory (Ar. A.).


\begin{thebibliography}{9}

\bibitem{vdm} H.J.A. Molegraaf {\it et al.}, 
Science {\bf 295} 2239 (2002).

\bibitem{mike_1} for a review see M. R. Norman and C. Pepin, 
\eprint{cond-mat/0302347} and references therein.


\bibitem{rosa}  J. Rossat-Mignod et al, Physica C {\bf 185-189}, 86 (1991)

\bibitem{fong} H. F. Fong  {\it et al}, \prb {\bf 61}, 14773 (2000); see also 
 C. Stock et al, cond-mat/0308186.

\bibitem{dai} P. Dai {\it et al.}, Science {\bf 284} 1344 (1999).

\bibitem{bi} H.F. Fong {\it et al.}, Nature {\bf 398}, 588 (1999).

\bibitem{keimer} H.F. He {\it et al.}, Science, {\bf 295}, 1045 (2002).

\bibitem{acs} Ar. Abanov, A.V. Chubukov, and J. Schmalian, 
Advances in Physics {\bf 52} 119 (2003); 
Journal of Electron Spectroscopy and Related Phenomena {\bf 117} 129 (2001).

\bibitem{ARPES_pdh} D.S. Dessau {\it et al.}, \prl {\bf 66}, 2160 (1991); 
M. R Norman et al, \prl {\bf 79}, 3506 (1997);
 A.V. Fedorov {\it et al.}, \prl {\bf 82}, 2179 (1999);  S.V. Borisenko \prl
{\bf 90}, 207001 (2003); A.D. Gromko {\it et al.}, \eprint{cond-mat/0205385}. 

\bibitem{kee} H.-Y. Kee, S. A. Kivelson, and G. Aeppli, \prl {\bf 88}, 257002
(2002).

\bibitem{reply} Ar. Abanov {\it et al.}, 
\prl {\bf 89} 177002 (2002).


\bibitem{basov} 
D. N. Basov {\it et al.}, Science, {\bf 283}, 49, 1999; 
A.F. Santander-Syro {\it et al.}, 
 Europhys. Lett. {\bf 62}, 568 (2003). 
For a review see D.N. Basov {\it et al.}, \rmp to appear.

\bibitem{ac_2} Ar. Abanov and A. Chubukov, \prl {\bf 88}, 217001 (2002).

\bibitem{comm3} This is similar to  a dirty  BCS superconductor, 
where the  optical integral changes sign  at 
$\omega \tau \approx  0.66$ -- Ar. Abanov, D. Basov, and A. Chubukov, 
\prb {\bf 68}, 024504 (2003).

\bibitem{homes} A similar conclusion follows from recent measurements on  $YBCO$ by C. Homes {\it et al.} \eprint{cond-mat/0303506}.

\bibitem{norman_pepin} J.E. Hirsch, Physica C {\bf 199}, 305 (1992); 
M. Norman {\it et al.}, 
\prb {\bf 61}, 14742 (2000); 
M. Norman and C. Pepin, \prb {\bf 66}, 100506 (2002).

\bibitem{rob}
 A. Chubukov and R. Haslinger,  Phys. Rev. B {\bf 67}, 
140504 (2003). 
\end{thebibliography}
\end{document}